\documentclass[preprint,12pt]{aastex}

\lefthead{Cunha et al.}
\righthead{Fluorine Abundances in Orion}

\received{2005 January 4}
\begin{document}

\title{Fluorine Abundances in the Orion Nebula Cluster} 

\author{Katia Cunha}
\affil{Observat\'orio Nacional, Rua General Jos\'e Cristino 77, 20921-400,
S\~ao Crist\'ov\~ao, Rio de Janeiro, Brazil; katia@on.br}

\author{Verne V. Smith}
\affil{National Optical Astronomy Observatory, P.O. Box 26732, Tucson,
AZ 85726, USA; vsmith@noao.edu}

\begin{abstract}

This is a pilot study using cool dwarfs as sources with which to
probe fluorine abundances via HF. This molecule is detected 
for the first time in young K-M dwarf members of an OB association.
The targets are three low-mass stars (JW22, JW163 and JW433)
belonging to the Orion Nebula Cluster. 
The target stellar parameters were derived to be T$_{\rm eff}$=3650K,
4250K, and 4400K, and log g =3.5, 3.4 and 3.6, with corresponding 
stellar masses of 0.4, 0.6 and 0.7M$_{\odot}$, for JW22, JW163 and JW433, 
respectively. 
Fluorine, oxygen, and carbon abundances were derived from the HF(1--0) R9 line
along with samples of OH  and CO vibration-rotation lines present in 
high-resolution infrared spectra 
observed with the Phoenix spectrograph on the Gemini South Telescope.
The fluorine and oxygen results obtained for these targets,
still in the pre-main-sequence stage of evolution,
agree well with the general trend defined for the Milky Way disk; the latter
being deduced from
observations of more evolved giant stars. In addition, 
the carbon and oxygen abundances obtained for the studied 
stars overlap results from previous studies of the more 
massive OB stars and FG dwarf members of the Orion Nebula Cluster.
We conclude from this agreement
that the fluorine abundances derived for the 
Orion K-M dwarfs (when there is no conspicuous evidence of disks) 
can be considered a good
representation of the current fluorine abundance value for the Milky Way disk.

\end{abstract}


\section{Introduction}

The chemical elements are produced via a variety
of nuclear and high-energy processes associated with stars
of various masses, cosmic rays, and the big bang itself.
One interesting process, whose importance within the
overall scheme of nucleosynthesis is not yet well-quantified, 
involves the inelastic scattering of neutrinos
off of nuclei and is referred to as neutrino nucleosynthesis,
or the $\nu$-process.
Neutrino nucleosynthesis was discussed by
Domogatsky, Eramzhyan, \& Nadyozhin (1978), who focussed mainly
on its role in the production of the lightest elements, such as
$^{2}$H, $^{3}$He, $^{7}$Li, or $^{11}$B.  Later, Woosley \& Haxton 
(1988) expanded on the role of the $\nu$-process in the production
of fluorine and subsequently, Woosley et al. (1990) published predicted yields
for a number of species, including Li and F.  
The neutrinos involved in the $\nu$-process 
are produced as a result of the gravitational collapse of a
stellar core to a neutron star during a supernova of Type II
(SN II).   
The $\nu$-process may result in significant synthesis of
certain low-abundance isotopes that lie one mass unit below
abundant nuclei, such as $^{12}$C giving rise to $^{11}$B,
or $^{20}$Ne resulting in $^{19}$F via neutrino-induced spallation.  
It is the association of $^{19}$F, the only stable isotope of
fluorine, with the $\nu$-process that makes its abundance of
interest as a potential probe of neutrino nucleosynthesis. 

In addition to the $\nu$-process, $^{19}$F is possibly produced
during He-burning thermal pulses on the asymptotic giant
branch (AGB), as first suggested by Forestini et al. (1992).  
Meynet \& Arnould (2000) also pointed out
that Wolf-Rayet stars, via the same sets of reactions as in
the AGB stars, may also produce significant amounts of $^{19}$F,
with the caveat that large dM/dt stellar winds must remove 
a significant amount of mass, exposing the
inner layers where fluorine is produced before it is burned away. 
Observations of fluorine abundances, spanning a range of
stellar metallicities and across a variety of stellar populations,
are needed in order to sort out relative contributions from AGB 
stars, Wolf-Rayet stars, or neutrino nucleosynthesis.  

In this paper, we use high-resolution IR spectra from Phoenix
on Gemini-South to derive $^{19}$F abundances (from HF), along
with $^{12}$C abundances (from CO), and
$^{16}$O abundances (from OH) in three low-mass members of the
Orion Nebula cluster.  
Carbon and oxygen provide comparison abundances to those
derived from fluorine.  
Both C and O have been studied
previously in Orion members by Cunha \& Lambert (1994); their
study analyzed hot O and B stars.  In addition, some oxygen
abundances in Orion-member F and G main-sequence and pre-main-sequence
stars have been published by Cunha, Smith, \& Lambert (1998).
These previous studies provide a background on which the abundances
derived from K and M type stars can be compared. Apart from the interest
in the carbon and oxygen abundances in young stellar systems, C and O
provide important comparisons for the fluorine abundances.    
The Orion Nebula cluster stars are very young and will reflect
the current Galactic F/O ratio in the disk.  Also, previous stellar
$^{19}$F abundances (derived from HF) have come only from red giants
(Jorissen, Smith, \& Lambert 1992; Cunha et al. 2003), while here we
undergo a pilot study to use cool dwarf stars as sources with which
to probe fluorine abundances. 

\section{Observations}


The target stars for this study are K and M dwarfs selected from various 
studies of the Orion Nebula cluster
by Hillenbrand (1997), Hillenbrand \& Hartmann (1998),
Hillenbrand et al. (1998), and Rhode, Herbst, \& Mathieu (2001).
In particular, the stars were selected for high membership
probability (P$\ge$ 93\%), to be cool enough for the HF molecular
line to be detectable (cooler than T$_{\rm eff}$ of 4500K), 
and to have relatively low values of projected rotational
velocities (vsin(i)$\le$15 km s$^{-1}$).  
The selected targets were such that two of them
(JW22 and JW433) do not show evidence of flux from a circumstellar disk,
while JW163 shows such evidence (primarily from near-IR
color excesses from Hillenbrand et al. 1998).  Our sample, although small, will allow us to
ascertain possible effects of disk emission on an abundance analysis
of the stellar photosphere in these young stars.

High-resolution infrared spectra of these low-mass stars 
were obtained with the Gemini South
Telescope plus the Phoenix spectrograph (Hinkle et al. 1998) during
observing runs in classical mode on 27-28 December 2002 and 
20 December 2004.
The observed data are single order echelle spectra with a 
resolution R=$\lambda$/$\Delta$$\lambda$=50,000, corresponding 
to a resolution element of $\sim$ 4-pixels.
In order to obtain two separate spectral regions, one around the H-band 
(at $\sim$ 15,500\AA, containing $^{16}$OH lines) and another around the
K-band (at $\sim$ 23,400\AA, containing the molecular HF(1--0) R9 line 
and numerous $^{12}$C$^{16}$O lines), two grating tilts were needed. JW22 was observed
only in the K-band region. 

The Orion targets are relatively bright in the near-infrared 
so that the exposure times
needed to obtain co-added spectra with signal-to-noise ratios in 
excess of $\sim$50--100 were not exceedingly long: for each grating tilt, 
four target exposures
of 15 minutes each were obtained. We adopted the strategy of 
varying the target position
on the slit (positions a-b-b-a) and subtracting one image from another
in order to remove dark current and fixed-pattern noise. 
In addition, for those observations corresponding to the 
spectral region of the K-band, 
we also observed hot stars at roughly the same air masses as the target 
stars.
These hot star spectra were needed in order to remove
telluric lines from the target star spectra.
All the obtained data were reduced to one-dimensional spectra using
standard IRAF routines. More details on the observations and 
reduction procedures adopted  can be found in Smith et al. (2004).

Phoenix spectra are shown in Figure 1 for two of the target stars, with
the two spectra shifted vertically in relative flux.  The spectra
plotted in this figure were obtained in the december 2002 Gemini run. The top panel
shows the spectra in the H band, while the bottom spectra are in the
K band.  Some of the prominent absorption lines are identified in the figure,
with the 15550\AA\ region showing a number of OH lines and the 23400\AA\ spectrum
containing numerous $^{12}$C$^{16}$O lines.

\section{Stellar Parameters}

The effective temperatures for the studied stars were derived from
their photometric colors (V-I) and (V-K) and by adopting the empirical
calibrations from Bessel, Castelli \& Plez (1998).
(V-I) and (V-K) colors are considered to be good temperature 
indicators for cool stars
and have little sensitivity to gravity for the effective
temperature range considered in this study.  The adopted photometric
magnitudes are presented in Table 1.
We obtained V-magnitudes and (V-I) colors for our targets from 
the work by Hillenbrand (1997) that studied the stellar 
population comprising the Orion Nebula cluster.
The target star K-magnitudes were obtained from the
Two Micron All Sky Survey (2MASS) database and corrected 
to the Bessel et al. system 
by means of the transformation equations presented in Carpenter (2001).
De-reddenned colors were
computed by adopting A$_{\rm v}$=0 for JW163, A$_{\rm v}$=1.01 mag 
for JW433, and
A$_{\rm v}$=0.8 for JW22 (all values are taken from Hillenbrand 1997). 

Surface gravities (log g's) for the target stars were calculated from
the following standard relation:

g/g$_{\odot}$ = (M/M$_{\odot}$) (T$_{eff}$/T$_{\odot}$)$^{4}$
(L/L$_{\odot}$)$^{-1}$.

The stellar luminosities were taken from Hillenbrand (1997;
log L/L$_{\odot}$= 0.22, 0.43, and -0.34 for JW163, JW433, and JW22,
respectively). These were derived
assuming a distance to the Orion Nebula of 470 pc. As discussed in
Hillenbrand (1997), the adopted stellar luminosities are estimated to 
have an uncertainty of 0.12 in log L/L$_{\odot}$. 
The effective temperatures adopted to derive log g's were 
those listed in Table 1.
Stellar masses could then be estimated from placement of the 
targets on an HR diagram 
and noting their position relative to 
evolutionary tracks computed by
D'Antona \& Mazzitelli (1994). 
Surface gravities were then computed for the sample stars. The derived masses and log g's
are presented in Table 1. From the uncertainties in the 
adopted stellar luminosities
and effective temperatures we estimate that our derived surface gravities
have a maximum uncertainty of $\sim$0.27 dex.

\section{Analysis}

LTE fluorine, as well as carbon and oxygen abundances were 
obtained from spectrum
synthesis using the synthesis code MOOG (Sneden, 1973).
The model atmospheres employed in the analysis were calculated with
the ATLAS9 code (Kurucz 1994) for solar metallicity, adopting a
mixing length l=1.25 H$_{p}$ and without convective overshooting.

\subsection{Fluorine, Carbon and Oxygen Abundances}

Fluorine abundances can be derived from the vibration-rotation
lines of HF present in the spectral region around the K-band. In this study, 
we used the HF(1--0) R9 line at 23357.661\AA. In this same IR spectral 
window where HF was observed,
several other molecular lines that correspond to 
$^{12}$C$^{16}$O 
first-overtone vibration rotation lines 
(3--1 vibration series) are also found and these can be used to obtain 
the carbon
abundances, given that the oxygen abundances are known from other 
transitions (such as OH).
Synthetic spectra were calculated across the entire spectral 
region of the observed
Phoenix spectra, or from $\sim$ 23340\AA\ to 23430 \AA.
The linelist adopted was the same as in our previous study of fluorine in
red-giants of the Large Magellanic Cloud (see Cunha et al. 2003).
The CO lines are typically used to derive the microturbulence in
cool giants, where they are much stronger (such as in Smith et al. 2002).
Although in the parameter range studied here, 
the CO lines become weaker, they are still sensitive to the microturbulence
and were used in order to estimate this parameter. The adopted
values for the stars vary between $\xi$=0.5 and 1.0 km/s (Table 1).

Figure 2 shows the synthetic and observed spectra for all three target
stars near the HF line.  Three synthetic spectra, each with a different
fluorine abundance, are plotted in each panel in order to illustrate
the sensitivity of the HF line to the fluorine abundance.  Just to the
blue of the HF(1--0) R9 line is a stellar photospheric line that is
probably due to H$_{2}$O.  This same unidentified line is clearly visible
(in the same approximate relative strength) in the solar sunspot spectrum
published by Wallace \& Livingston (1992).  They identify a number of
other H$_{2}$O lines in this spectral region that have similar strength.
Water is not included in the synthesis linelist here, and the offending
line blends just a small fraction of the blue edge of the HF line and
does not affect significantly the derived $^{19}$F abundance. If triatomic 
H$_{2}$O is responsible for this line, then it should be much weaker in giant
stars of similar effective temperature, and this
can be seen in the spectal atlas of Arcturus (Hinkle, Wallace, \&
Livingston 1995); although having a similar effective temperature to
JW163 or JW433, this feature is not detectable in Arcturus, even though
the HF line is clearly visible.  Indeed, a tentative identification
has been provided by Hinkle (2004) as 
H$_{2}$O (021)[14,5,9]--(010)[13,11,2]
at an air wavelength of about 23356.4\AA.

Oxygen abundances can be measured from first-overtone 
vibration-rotation
OH lines present in the H-band region. Three isolated OH lines at 15560,
15569 and 15572\AA\
were synthesized to derive oxygen abundances. 
The studied OH lines and adopted linelist were the same as in 
Smith et al. (2002).  
These lines are expected
to be reliable oxygen
abundance indicators and relatively free from non-LTE effects but to
suffer modestly ($\sim$ 0.1 dex) from 3d-effects in solar-type stars 
(Asplund 2005). However, at
the relatively low effective temperatures of the stars studied here, 
the oxygen abundance differences
caused by the adoption of one-dimensional model atmospheres, 
instead of three-dimensional
model atmospheres, are expected to be smaller 
than at higher effective temperatures (around 6000K; Nissen et al. 2002).

Overall, the derived abundances in JW163, relative to JW433 and JW22, are
systematically lower for the studied elements, with $^{12}$C 0.30-0.40 dex
lower, $^{16}$O 0.25 dex lower, and $^{19}$F 0.10 dex lower.  Although
it is possible that this difference reflects an underlying abundance
spread in the Orion Nebula cluster, it is more likely that
the presence of a more substantial circumstellar disk in JW163
results in the photospheric absorption lines being filled in
by disk emission.  It is worth noting that in the (V--I)-(V--K)
plane as shown in Figure 11 of Bessell et al. (1998), both JW433 
and JW22 fall
right on the sequence defined by normal dwarf stars, while JW163
is displaced several tenths of a magntiude towards redder colors from
the standard dwarf relation.  In JW433 and JW22, with little if any
measurable disk presence, the abundances are near-solar:
A($^{12}$C)=8.30, A($^{16}$O)=8.80, and A($^{19}$F)=4.65 for JW433,
while for JW22 the abundances are A($^{12}$C)=8.40 and 
A($^{19}$F)=4.61.  Moreover, as will be discussed in Section 5,
the carbon (for JW22 and JW433) and oxygen (for JW433)
abundances overlap nicely
with those values derived from the much hotter B-star members of
Orion Id (the sub-association to which the Orion Nebula cluster
belongs) by Cunha \& Lambert (1994).  The results from JW163, along
with JW22 and JW433 illustrate the usefullness of using these cool
dwarfs for abundance studies, even those still not on the main sequence,
provided that targets are selected carefully, taking into account
their circumstellar environments.  

\subsection{Abundance Uncertainties}

The final abundances derived for carbon, oxygen and fluorine are presented in Table 1.
The expected uncertainties in the derived effective temperatures and surface gravities
are 100K and 0.3 dex, respectively. The errors in the microturbulence parameters
are  estimated to be around 0.5 km/s.
The sensitivity of the derived
C, O and F abundances to changes in temperature, gravity and microturbulence 
for two target stars (the coolest and the hottest in our sample)
are presented in Table 2; the total errors ($\Delta$A) in the derived
abundances in each case are listed in column 6.
Overall, the studied transitions of HF, CO and OH are weak and relatively
insensitive to the microturbulence; the exception being the CO lines that are
moderately strong at T$_{\rm eff}$'s around 3650K. The fluorine abundances are mostly
sensitive to effective temperature, but also, to a lesser extent, to the surface
gravity. Concerning OH, the abundance errors are quite insensitive to changes
in surface gravity and microturbulence. The total abundance errors for oxygen are dominated
by T$_{\rm eff}$ uncertainties. (OH abundances were not calculated for JW22 in this study.)
The uncertainties in the derived carbon abundances are shown to have different sensitivities
at the "cool" and "hot" regimes studied here. These are more uncertain around 3600K
when compared to 4500K.

\section{Discussion}

Our larger goal is to investigate the behavior of
fluorine abundances with metallicity, which is taken here to be represented by
the element oxygen, as this element is
virtually a pure product of nucleosynthesis in massive stars that end
their evolution as core-collapse SN II.
In particular, in this study we focus on
the fluorine and oxygen contents of the pre-main-sequence stellar members
of the Orion Nebula Cluster.

As mentioned in the introduction carbon and oxygen abundances derived from
molecular transitions (CO and OH) can be used as benchmarks with which
to gauge the fluorine abundances obtained from HF.
A comparison of carbon and oxygen abundances derived here for JW22
and JW433 with abundances from Cunha \& Lambert (1994) and
Cunha et al. (1998) is in order; we ignore JW163 in this comparison
as there is evidence that the absorption lines may be affected
measurably by emission from a circumstellar disk.  The Orion
Association is divided into four subgroups, labelled Ia, Ib, Ic, and
Id. The subgroups are distributed roughly by age, with Ia being the
oldest, Id the youngest, and an age difference of about 10Myr.
The Orion Nebula stars analyzed in this study are members of Id.  The abundances
in the two stars are A(C)=8.40 for JW22, and A(C)=8.30 and A(O)=8.80
for JW433.  For comparison, Cunha \& Lambert (1994) studied three OB
stars in Id and their average (NLTE) abundances are A(C)=8.25 and
A(O)=8.85: the agreement between two K-M dwarfs and three O-B stars
is excellent.  In addition, Cunha et al. (1998) have one F-dwarf that
is a Id member and its oxygen abundance is A(O)=8.86: again, excellent
agreement.  Oxygen and carbon abundances in the OB stars come from
analyses of O II and C II lines, while oxygen in the F dwarf
comes from an analysis of O I lines.  The K-M dwarfs use molecular
OH and CO lines.  This abundance comparison across a large range
in T$_{\rm eff}$ utilizing a variety of spectroscopic signatures
provides evidence that the abundance analysis of the low-mass stars
studied here is robust.  We proceed to a discussion of the fluorine
abundances.        

Historically,  
$^{19}$F abundances were derived initially in a sample of
galactic K, M, S, and C red-giants, with metallicities mostly
near solar, by Jorissen et al. (1992); this study
was restricted to fairly bright stars in the IR.  More recently,
access to high spectral resolution in the IR using the Phoenix
spectrograph on the Gemini-South 8.1m telescope has allowed 
fluorine abundances to
be measured in fainter samples of cool stars, such as in the
Large Magellanic Cloud (LMC) or in the globular cluster $\omega$
Centauri (Cunha et al. 2003).  Renda et al. (2004) used the abovementioned
$^{19}$F abundances as comparisons to chemical evolution models.
They conclude that both Wolf-Rayet and AGB stars are significant
sources of $^{19}$F, with the Wolf-Rayet stars dominating at high
metallicities. At low metallicities (less than roughly -0.8 dex)
synthesis via the $\nu$ process in SN II dominates fluorine
production. 
Renda et al. (2004) were the first to show that all three $^{19}$F 
sources can be significant contributors,  but with differing relative
contributions depending on the metallicity and star formation
history of the stellar population.
In addition to abundances derived from stellar HF lines, Federman
et al. (2005) have obtained detections of interstellar F I lines
(at 955\AA) along two lines-of-sight into the Cep OB2 association
using the Far Ultraviolet
Spectroscopic Explorer (FUSE). Their derived column densities of
F and O indicate a near-solar F/O ratio, with some small depletion
effects.   

The Orion stars analyzed here are of particular interest 
as they represent a different type of star with which
to probe fluorine when compared to previous studies since
Jorissen et al. (1992) sampled only evolved stars.
Our work adds three stars of higher surface gravity to
the sample of near-solar metallicity field stars and
their $^{19}$F and $^{16}$O abundances are shown in
Figure 3 in the form of [F/O] versus A(O).
The point represented by JW163 should be viewed with caution
as we have noted already that its abundance results have probably been
influenced by the presence of a circunstellar disk. Although both O and F
could have been affected about the same with the result that the ratio
of F/O is changed relatively less.
Also, we adopt the same oxygen abundance for JW22 as we derived
for JW433 because the 15550\AA\ region with the OH lines was
not observed for JW22; however its carbon abundance is the
same, within the errors, as found for JW433 and there is no
evidence that the Orion stars exhibit large variations in their
total C/O values. 
The Jorissen et al. (1992) K and M giants are also
plotted (with the abundances in these stars re-derived
by Cunha et al. (2003)); the solar values in Figure 3 
are taken to be A(O)= 8.69 and
A(F)= 4.55.  Note that the Orion
dwarf abundances overlap very nicely with the Jorissen et al.
giants.  

The stellar abundances that are plotted in Figure 3 represent
all the available fluorine abundances that have been measured for members of the
Milky Way so far (excluding self-polluted AGB stars).
Abundances for LMC red giants from
Cunha et al. (2003) are not shown here, as the star formation
history and corresponding chemical evolution in the LMC may be
quite different from the Milky Way.  The solid curve in Figure 3
is a chemical evolution model from Renda et al. (2004) for the
Milky Way.  This particular curve represents their MWc model that
includes $^{19}$F yields from all three stellar production sites:
SN II with the $\nu$-process, Wolf-Rayet stars, and AGB stars.
The Orion Nebula cluster F and O abundances help anchor the Milky
Way values at high metallicity, which are fit nicely by the Renda
et al. (2004) model with all three sources.  The low values of
[F/O] in the $\omega$ Cen stars were discussed by both Cunha et al.
(2003) and Renda et al. (2004) in the context of the unusual star
formation history and chemical evolution within this self-enriched
stellar system that has been captured by the Galaxy. 

\section{Conclusions}

We detect the HF molecular line at 23400\AA\ in K-M dwarfs and 
present the first fluorine
abundance measurements in young stars still in 
the pre-main sequence phase of evolution. These targets are members
of the Orion
association, in particular from the
Orion Nebula cluster, with masses around 0.4-0.8 M$_{\odot}$. 
Our results for JW433 and JW22, which do not show substantial emission
from a circumstellar disk, reveal abundances of C and O that are
in excellent agreement with those that are found in B-type stellar
members of Orion Id (Cunha \& Lambert 1994).  This agreement
bolsters confidence in the quantitative spectroscopic abundance
analyses techniques used in the hotter B stars, as well as the
cool K and M dwarfs. Moreover,    
JW433 and JW22 also display abundance patterns in
agreement with the general behavior of fluorine versus oxygen for the
Milky Way
disk that has been established from stars in a distinctly
different evolutionary state.
All previous F abundances had been derived from studies of
red giants. The overlap in these abundances provides 
an important confirmation for the 
relation of the fluorine abundance with oxygen in the Disk.

In the future, larger samples of K and M dwarfs, 
as well as L and T dwarfs, can
be used to measure fluorine and oxygen abundances both in the
field and in the nearer star-forming regions.

Based on observations obtained at the Gemini Observatory,
which is operated by the Association
of Universities for Research in Astronomy, Inc., under a cooperative
agreement with the NSF on behalf of the Gemini partnership: the National
Science Foundation (United States), the Particle Physics and Astronomy
Research Council (United Kingdom), the National Research Council (Canada),
CONICYT (Chile), the Australian Research Council (Australia),
CNPq (Brazil), and CONICRT (Argentina), as program GS-2002B-Q-2.
This work is also supported in part by the National Science Foundation through
AST03-07534 (VVS) and NASA through NAG5-9213 (VVS).

\clearpage
\begin{bibliography}{}
\noindent
Allende Prieto, C., Lambert, D. L., \& Asplund, M. 2000, ApJ, 573, L137\\
\smallskip

Asplund, M. 2005, to appear in Cosmic Abundances as Records of Stellar
Evolution and Nucleosynthesis, ASP Conference Series, eds. F. N. 
Bash \& T. G. Barnes (San Francisco:ASP)\\
\smallskip

Bessel, M.S., Castelli, F. \& Plez, B. 1998, A\&A, 333, 231\\
\smallskip

Carpenter, J. M. 2001, ApJ, 121, 2851\\
\smallskip

Cunha, K., Smith, V.V., Lambert, D.L. \& Hinkle, K.H. 2003, AJ, 126, 1305\\
\smallskip

D'Antona, F. \& Mazzitelli, I., 1994, ApJS, 90, 467\\
\smallskip

Domogatskii, G. V., Eramzhian, R. A., \& Nadezhin, D. K. 1978, Ap.\& Space
Sci., 58, 273\\
\smallskip

Federman, S. R., Sheffer, Y., Lambert, D. L., \& Smith, V. V. 2005,
ApJ, in press (astro-ph/0410362)\\
\smallskip

Forestini, M., Goriely, S., Jorissen, A., \& Arnould, M. 1992,
A\&A, 261, 157\\
\smallskip

Hillenbrand, L. A. 1997, AJ, 1733, 1768\\
\smallskip

Hillenbrand, L. A., \& Hartmann, L. W. 1998, ApJ, 492, 540\\
\smallskip

Hillenbrand, L. A., Strom, S. E., Calvet, N., Merrill, K.M., 
Gatley, I., Makidon, R.B., Meyer, M.R., \& Skrutskie, M.F., 1998, AJ, 
116, 1816\\
\smallskip

Hinkle, K. H. 2004, private communication\\
\smallskip

Hinkle, K. H., Wallace, L., \& Livingston, W. 1995, Infrared Atlas
of the Arcturus Spectrum, 0.9--5.3$\mu$m (San Francisco:ASP)
\smallskip

Hinkle, K. H., Cuberly, R., Gaughan, N., Heynssens, J.,
Joyce, R., Ridgway, S., Schmitt, P., \& Simmons, J. E. 1998,
Proc. SPIE, 3354, 810\\
\smallskip

Jorissen, A., Smith, V.V., \& Lambert, D.L., A\&A, 261, 164\\
\smallskip

Kurucz, R. L. 1994, CD-ROM 19 (Cambridge:SAO)\\
\smallskip

Meynet, G., \& Arnould, M. 2000, A\&A, 355, 176\\ 
\smallskip

Nissen, P.E., Primas, F., Asplund, M., \& Lambert, D.L. 2002, A\&A 390, 235\\
\smallskip

Renda, A. et al. 2004 , MNRAS, 354, 575\\
\smallskip

Rhode, K.L., Herbst, W., \& Mathieu, R.D. 2001, AJ, 122, 3258\\
\smallskip

Sneden, C. 1973, ApJ, 184, 839\\
\smallskip

Smith, V.V., Hinkle, K.H., Cunha, K., Plez, B., Lambert, D.L., 
Pilachowski, C.A., Barbuy, B., Melendez, J., Balachandran, S., 
Bessel, M.S., Geisler, D.P., Hesser, J.E., \& Winge, C., 2002, 
AJ, 124, 3241\\ 
\smallskip

Wallace, L., \& Livingston, W. 1992, An Atlas of a Dark Sunspot Umbral
Spectrum from 1970 to 8640 cm$^{-1}$ (1.16 to 5.1$\mu$m), N.S.O.
Technical Report \#92-001\\
\smallskip 

Woosley, S.E.,  Haxton, W.C., 1988, Nature, 334, 45\\
\smallskip

Woosley, S.E., Hartmann, D.H., Hoffman, R.D., Haxton, W.C., 1990, ApJ, 356,
272\\
\smallskip

\end{bibliography}

\clearpage

\begin{deluxetable}{ cccccccccccc }  
\setcounter{table}{0}  
\tablewidth{400pt}  
\tablecaption{Stellar Parameters and Abundances}

\tablehead{  Star &   
\multicolumn{1}{c} {V$_{0}$} &
\multicolumn{1}{c} {I$_{0}$} &
\multicolumn{1}{c} {K$_{0}$} &
\multicolumn{1}{c} {T$_{\rm eff}$} &
\multicolumn{1}{c} {Log g} &
\multicolumn{1}{c} {$\xi$} &
\multicolumn{1}{c} {M/M$_{\odot}$} &
\multicolumn{1}{c} {A($^{12}$C)} &
\multicolumn{1}{c} {A($^{16}$O)} &
\multicolumn{1}{c} {A($^{19}$F)}
} 
\startdata 
 JW 22  & 15.5 & 13.4 & 11.3 & 3650 & 3.6 & 1.0 & 0.4 &  8.40 &   -- & 4.61 \\ 
 JW 163 & 13.1 & 11.8 & 10.2 & 4250 & 3.5 & 0.5 & 0.6 &  8.00 & 8.55 & 4.53 \\
 JW 433 & 12.5 & 11.3 & 9.6  & 4400 & 3.4 & 0.8 & 0.7 &  8.30 & 8.80 & 4.65 \\
\enddata
\tablecomments{A(X)= log[n(X)/n(H)] + 12.  
}
\end{deluxetable}

\clearpage

\begin{deluxetable}{ cccccc }  
\setcounter{table}{1}  
\tablewidth{400pt}  
\tablecaption{Abundance Uncertainties}

\tablehead{  Star &   
\multicolumn{1}{c} { Element} &
\multicolumn{1}{c} { $\Delta$T$_{\rm eff}$(+100K)} &
\multicolumn{1}{c} {$\Delta$Log g(+0.3 dex)} &
\multicolumn{1}{c} {$\Delta$$\xi$(+0.5 km/s)} & 
\multicolumn{1}{c} {$\Delta$A}
} 
\startdata 
 JW 22  & $\Delta$C & 0.08 & 0.0  & 0.10 & 0.13 \\ 
 JW 22  & $\Delta$F & 0.08 & 0.05 & 0.0  & 0.09 \\
 JW 433 & $\Delta$C & 0.15 & 0.15 & 0.0  & 0.21 \\ 
 JW 433 & $\Delta$O & 0.10 & 0.02 & 0.0  & 0.10 \\
 JW 433 & $\Delta$F & 0.13 & 0.05 & 0.0  & 0.14 \\
\enddata
\end{deluxetable}

\clearpage

\figcaption[fig1.ps]{Observed Phoenix spectra for the target stars
JW163 and JW433 centered at 15545 \AA (top panel) and 23390 \AA.
The spectra have been separated vertically, while prominent absorption 
lines are indicated.
\label{fig1}}

\figcaption[fig2.ps]{Observed and synthetic spectra for the three target
stars in the HF region. The best-fit fluorine abundances (represented
by solid lines) are those presented in Table 1. The synthetic
spectra represented by the dashed lines were calculated with 
$\pm$ 0.2dex changes in the $^{19}$F abundance.  Note the blending
feature just blueward of the HF(1-0) R9 line, which is identified
tentatively here as H$_{2}$O.
\label{fig2}}

\figcaption[fig3.ps]{Values of [F/O] are plotted versus oxygen
abundances for field K and M giants from Cunha et al. (2003) and 
the three Orion Nebular cluster stars. The solar symbol is shown
and the solid line illustrates the predictions from one chemical
evolution model by Renda et al. (2004) that includes all three 
suggested sources for $^{19}$F production.  Note that the Orion
Nebula stars fall within the scatter defined by the red giants
of the Milky Way field.
\label{fig3}}

\clearpage

\begin{figure}
\plotone{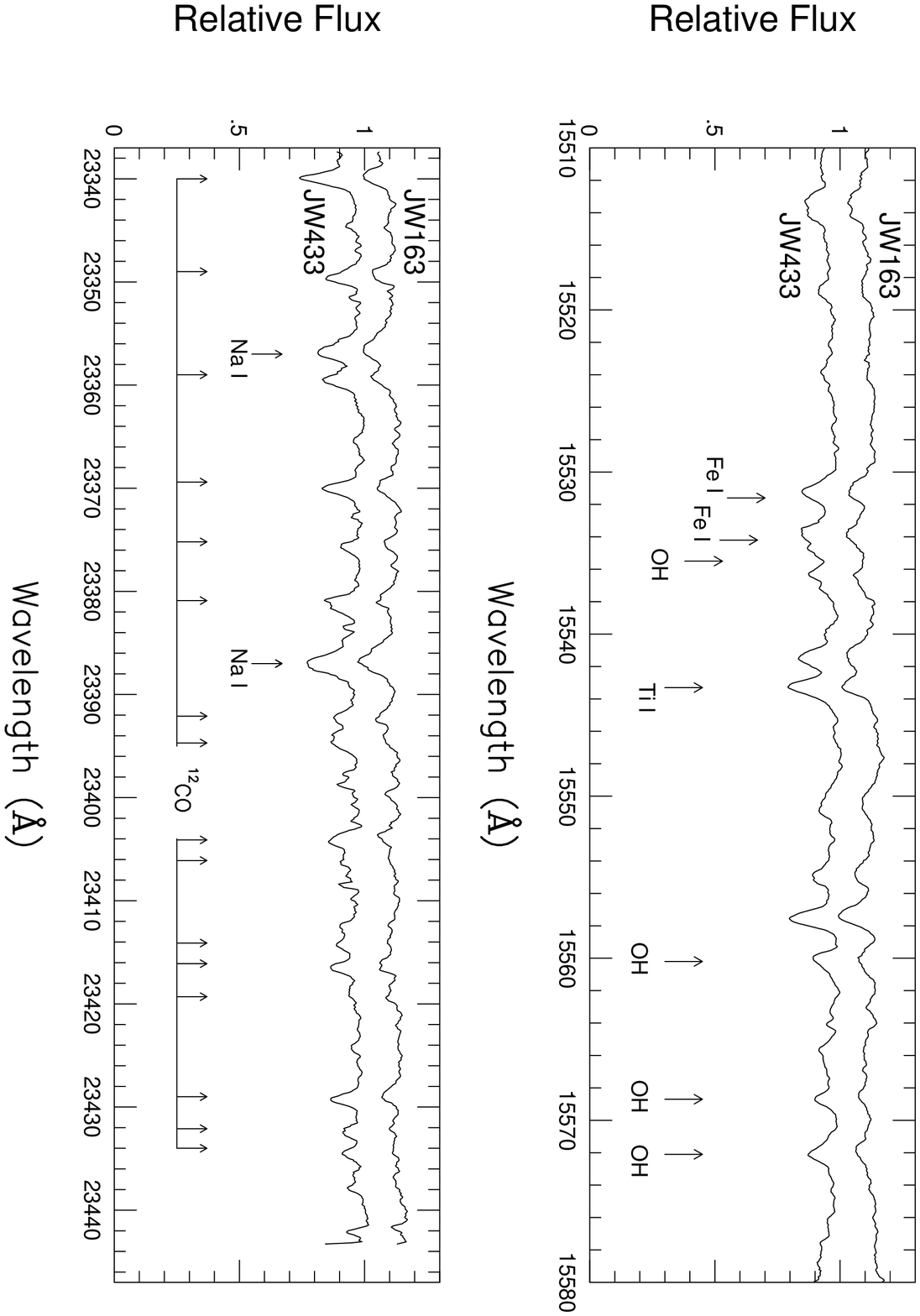}
\end{figure}

\clearpage

\begin{figure}
\plotone{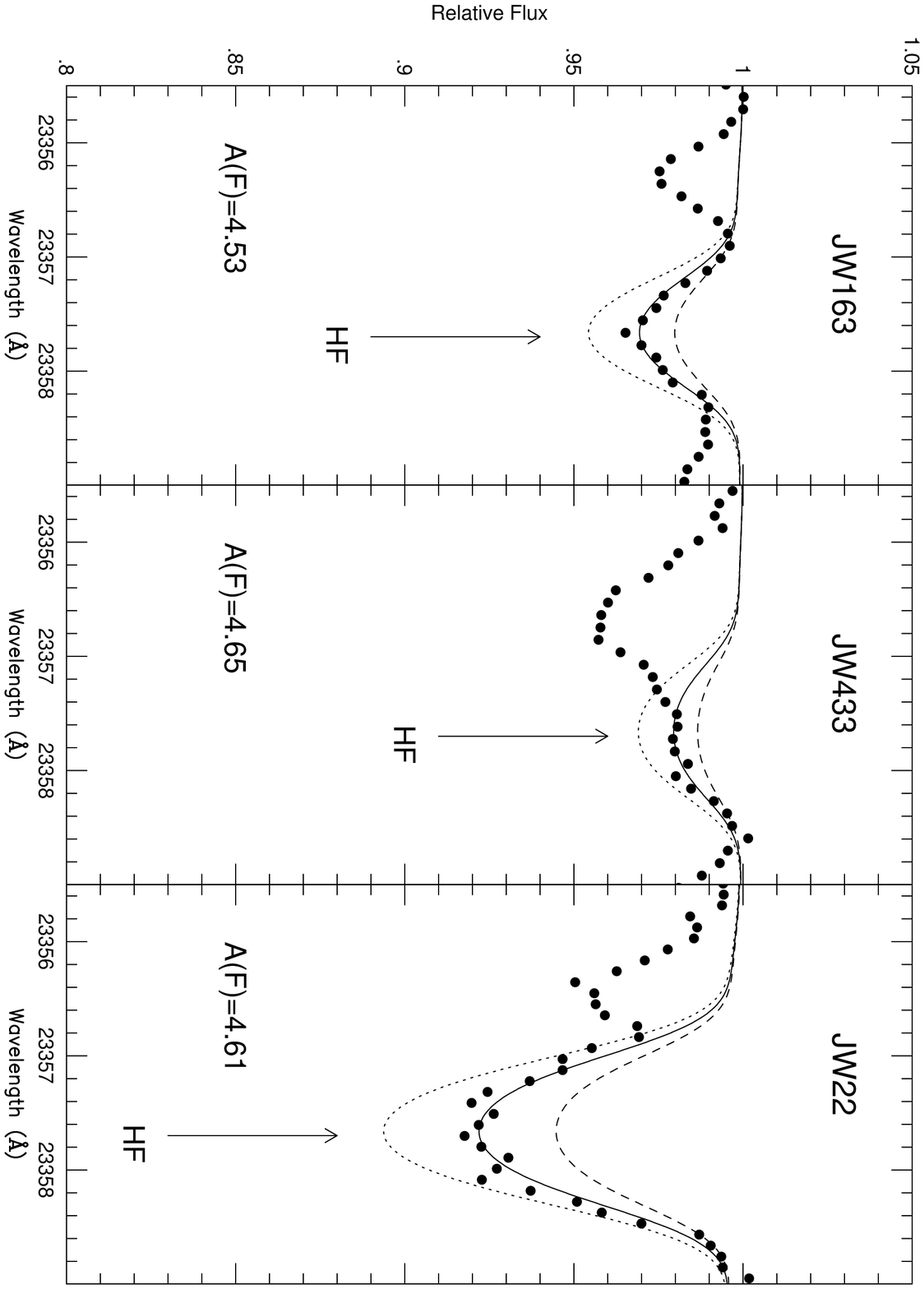}
\end{figure}

\clearpage

\begin{figure}
\plotone{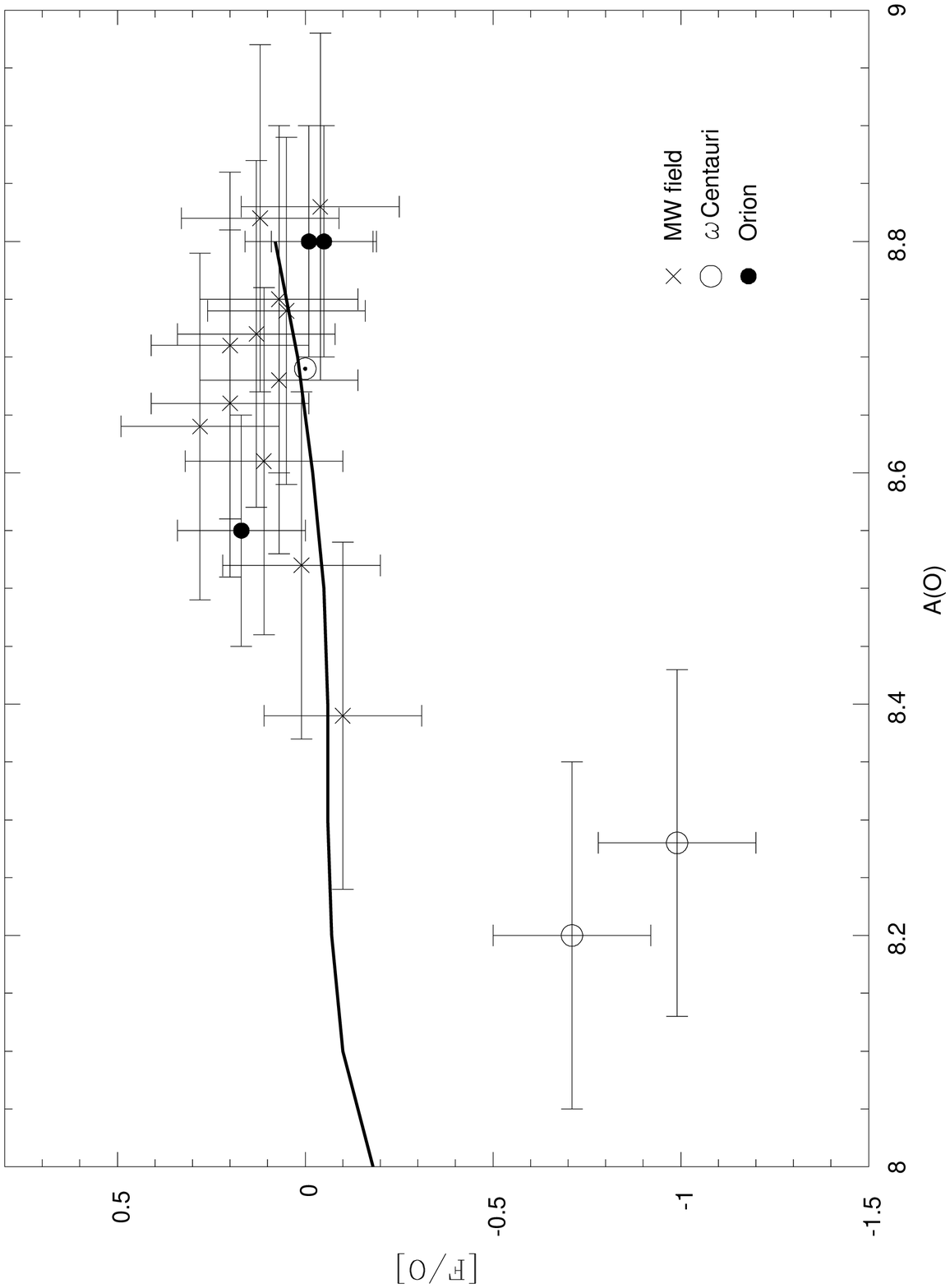}
\end{figure}
\end{document}